\documentstyle[12pt]{article}

\begin{document}

\setlength{\baselineskip}{3.0ex} \baselineskip 17pt

\begin{titlepage}
\addtocounter{page}{1}      

\vspace{2.5em}  

{\Large\bf                                                           
\begin{center}     
{\large\bf     
Thermodynamic Equations of State for Dirac and Majorana Fermions }
\end{center} }
\vspace{1.5em}

{\large\bf
\centerline{Y. N. Srivastava and A. Widom}
\centerline{Physics Department, Northeastern University, Boston MA 02115}
\centerline{and}
\centerline{Physics Department \& INFN, University of Perugia, Perugia, Italy}
\vspace{0.4cm}

\centerline{and}

\vspace{0.4cm}

\centerline{J. Swain}\footnote{email: swain@neuhp1.physics.neu.edu; fax: (617) 373-2943}
\centerline{Physics Department, Northeastern University, Boston MA 02115}
}

\vskip .4in
\samepage
\vspace{3em}
\begin{center}
{\bf \large Abstract}
\end{center}       

\noindent

The thermodynamic equations of state for Majorana and Dirac fermions
are quite different even in the limit of zero mass. The corresponding
equations are derived from general principles, and then applied
to neutrinos. The nature of the approach to equilibrium is explored,
along with the subtleties of low or zero mass neutrinos interacting 
via purely left-handed currents. Possible applications to cosmology, 
supernovae, and supersymmetry are discussed.

\end{titlepage}

\pagebreak

\section{Introduction}

\medskip

Fermions are usually described with 4-component complex spinors which
satisfy the Dirac equation. The four complex components describe two spin
states for a particle and two spin states for the antiparticle. These are
the usual ``Dirac'' fermions, such as electrons.

In the event that the fermions have no charges or electric or magnetic
dipole moments, the possibility exists that the particle and antiparticle
states are identical, just as the photon or neutral pion states and their
antiparticle states are identical. This constraint applied to a Dirac
fermion yields a 2-complex component (or 4-real component) spinor
representing a particle which is identical to its antiparticle: a
``Majorana'' fermion.

Perhaps surprisingly, it remains in many workers minds an open question as
to whether or not neutrinos are their own antiparticles. The argument has
been that in the Standard Model with only left-handed couplings, a putative
right-handed neutrino would not (beyond gravity) interact. It is then
conceivable that the two observed neutrino states, one called a left-handed
neutrino and the other called a right-handed antineutrino are actually two
helicity states of the {\it same} Majorana particle. The antineutrino of the
Dirac theory here appears as the right neutrino of the Majorana theory and
observed neutrinos might then be Majorana. 

In the absence of right-handed couplings, and with massless Dirac neutrinos
with no amplitudes connecting left and right-handed components, the right
handed neutrinos would appear to be not detectable. For neutrinos of small
masses $m$, amplitudes involving the right-handed components are often
suppressed by factors of order $m/M_W$, where $M_W$ is the W-boson mass,
characteristic of the weak scale. This leads to the often-quoted ``practical
Majorana-Dirac confusion theorems'' \cite{ref:Kayser}, essentially stating
that if one does experiments on a left-handed neutrino, it gets harder and
harder to tell if it's Majorana or Dirac as its mass tends to zero.

This note addresses a different situation, where we consider systems of more
than one neutrino. Our purpose here is to exhibit thermodynamic equations of
state for an ideal gas of Dirac neutrinos and for an ideal gas of Majorana
neutrinos. Thermodynamic equations of state have some importance for
astrophysical and cosmological models. Before considering thermodynamic
systems with large numbers of particles let us first consider the
(anti-symmetry) properties of wave functions for systems of more than one
neutrino, and apply them to the simplest multineutrino process: production
of pairs \cite{ref:AlYogi}.

\section{Anti-symmetry for Dirac and Majorana Fermions}

The central point is that all Dirac wave functions of momenta ${\bf p}_i$
and spins $s_i$ for particles and antiparticles,

\begin{equation}
\Psi_{Dirac}=\Psi_{Dirac}({\bf p}_1,s_1,{\bf p}_2,s_2,...; \bar{{\bf p}}_1,%
\bar{s}_1,\bar{{\bf p}}_2,\bar{s}_2,... ), 
\end{equation}

\noindent are anti-symmetric with respect to particle exchange $({\bf p}%
_i,s_i)\leftrightarrow({\bf p}_j,s_j)$, $(i\ne j)$, and with respect to
antiparticle exchange $(\bar{{\bf p}}_i,\bar{s}_i)\leftrightarrow (\bar{%
{\bf p}}_j,\bar{s}_j)$, $(i\ne j)$. However, there is no particular enforced
exchange symmetry between a particle exchanged with an antiparticle since
these two objects have different fermion number.

For the Majorana case, there is no distinction between a particle and an
antiparticle, and no conserved fermion number.

For a Majorana wave function

\begin{equation}
\label{eq:majsym}\Psi_{Majorana}=\Psi_{Majorana}({\bf p}_1,s_1,{\bf p}%
_2,s_2,...) 
\end{equation}

\noindent there is an anti-symmetry under the exchange $({\bf p}%
_i,s_i)\leftrightarrow({\bf p}_j,s_j)$, $(i\ne j)$, for all neutrinos.

\section{Pair Production of Neutrinos}

The exchange symmetry of wave functions is above and beyond any explicit
(standard or otherwise) model assumptions about interactions \cite{ref:LL}.
Exchange anti-symmetry has some profound implications. As an example of the
consequences of anti-symmetric wave functions, contrast the following
neutrino reaction for the two cases of interest, 
\begin{equation}
e^-+e^+\to \nu_1+\bar{\nu}_1\ \ \ (Dirac),\ \ \ \ e^-+e^+\to \nu_1+\nu_2\ \
\ (Majorana). 
\end{equation}

\noindent For the Dirac case, in the limit of zero mass and in the center of
mass reference frame, there is a parity violating forward-backward asymmetry
in the scattering angle $\theta $ of the differential cross section

\begin{equation}
d\sigma (e^-e^+\to \nu_1\bar{\nu}_1)= |f_{Dirac}^{\bar{\nu} \nu}(\theta )|^2
d\Omega \ne |f_{Dirac}^{\bar{\nu} \nu}(\pi -\theta )|^2d\Omega . 
\end{equation}

For the Majorana case, in the limit of zero mass and in the center of mass
reference frame, the opposite helicities of the two neutrinos would require
that the two neutrino spins are parallel (say at angle $\theta $ the
parallel spins $\nearrow \nearrow $). Quantum mechanics for identical
fermions with parallel spins dictates that

\begin{equation}
f_{Majorana}^{\nearrow \nearrow }(\theta ) =f(\theta )-f(\pi -\theta ), 
\end{equation}

\noindent so no Majorana forward-backward asymmetry can exist :

\begin{equation}
d\sigma (e^-e^+\to \nu_1\nu_2)= |f_{Majorana}^{\nearrow \nearrow }(\theta
)|^2 d\Omega =|f_{Majorana}^{\nearrow \nearrow }(\pi -\theta )|^2d\Omega . 
\end{equation}

This process evades the ``practical confusion theorem'' by having more than
one neutrino in the problem, and quantum mechanics makes a sharp
distinction, independent of mass, in the production of distinct or identical
particles. In the following sections we will study the thermodynamics of
systems with many Dirac or Majorana neutrinos.

\section{Equations of State for Gases of Fermions}

In the following sections we will call the Dirac of Majorana fermions
``neutrinos'' without loss of generality, and with the understanding that
the arguments apply to any fermions.

\subsection{Majorana Neutrino Thermal Equations of State}

Before calculating the equation of state for an ideal gas of Majorana
neutrinos, let us first consider an ideal gas of photons. For the photon
case, the following holds true: (i) The photon wave function

\begin{equation}
\Psi_{photon}=\Psi_{photon}({\bf K}_1,S_1,{\bf K}_2,S_2,...) 
\end{equation}

\noindent is symmetric under the exchange $({\bf K}_i,S_i)\leftrightarrow(%
{\bf K}_j,S_j)$, $(i\ne j)$; (ii) The distinction between particle and
antiparticle for the photon is null and void; (iii) Photon number is not
conserved. From a statistical mechanical viewpoint, the above requires that
only zero chemical potential needs to be considered for the photon. The
grand canonical ensemble pressure is then

\begin{equation}
\label{eq:photonpress}P_{photon}=-k_BT\Big\{2\int \Big({\frac{d^3K}{(2\pi)^3}%
}\Big) 
ln\big(1-e^{-\hbar cK/k_BT}\big)\Big\}. 
\end{equation}

Equation \ref{eq:photonpress} yields the usual result,

\begin{equation}
P_{photon}(T)=\Big({\frac{\pi^2 \hbar c}{45}}\Big)
\Big({\frac{k_BT}{\hbar c}}\Big)^4 , 
\end{equation}

\noindent from which all other thermal properties follow.

For Majorana fermions the following holds true: (i) The neutrino wave
function in Eq.(\ref{eq:majsym}) is anti-symmetric; (ii) The distinction
between particle and antiparticle for the Majorana neutrino is null and
void; (iii) Fermion number is not conserved. From a statistical mechanical
viewpoint, the above requires that we set the chemical potential to zero for
the Majorana neutrino. The grand canonical ensemble pressure is then

\begin{equation}
\label{eq:pmaj}P_{Majorana}=k_BT\Big\{2\int \Big({\frac{d^3p}{(2\pi \hbar )^3%
}}\Big) 
ln\big(1+e^{-\varepsilon /k_BT}\big)\Big\},
\end{equation}

\noindent where

\begin{equation}
\label{eq:relativity}\varepsilon =\sqrt{c^2p^2+m^2c^4}.
\end{equation}

>From Eqs.(\ref{eq:pmaj}) and (\ref{eq:relativity}) it follows that

\begin{equation}
P_{Majorana}=\hbar c \Big({\frac{k_BT}{\hbar c}}\Big)^4 {\cal F}_{Majorana}%
\Big({\frac{mc^2}{k_BT}}\Big), 
\end{equation}

\noindent where

\begin{equation}
{\cal F}_{Majorana}(x)=\Big({\frac{1}{\pi^2}}\Big)\int_x^\infty \big(y\sqrt{%
y^2-x^2}\big)ln\big(1+e^{-y}\big)dy. 
\end{equation}

In the limit of zero mass, 
\begin{equation}
\label{eq:majoranapzero}\lim _{m\to 0}P_{Majorana}(T)=\Big({\frac{7\pi
^2\hbar c}{360}}\Big)
\Big({\frac{k_BT}{\hbar c}}\Big)^4, 
\end{equation}

\noindent where ${\cal F}_{Majorana}(0)=(7\pi^2/360)$ has been employed.
Thus, in the zero mass limit and with zero chemical potential, $P\sim (\hbar
c/\lambda^4)$ where the thermal wave length $\lambda \sim (\hbar c/k_BT)$.
The above results from dimensional analysis hold true for both zero mass
photons and zero mass Majorana neutrinos.

\subsection{Dirac Neutrino Thermal Equations of State}

For the case in which fermion number $N$ is conserved, one may choose the
grand canonical partition function with a chemical potential $\mu $ to be

\begin{equation}
Z_{Grand}=Tr\Big\{e^{-(H-\mu N)/k_BT}\Big\}, 
\end{equation}

The grand canonical pressure (in the thermodynamic limit of infinite volume $%
V$),

\begin{equation}
\label{eq:grandpdirac}P(T,\mu )=\lim_{V\to \infty}\Big({\frac{k_BT}{V}}%
\Big)
lnZ_{Grand}(T,\mu,V), 
\end{equation}

\noindent provides a complete determination of thermodynamic equations of
state. The thermodynamic law reads

\begin{equation}
\label{eq:dp}dP=sdT+nd\mu, 
\end{equation}

\noindent where $s$ and $n$ denote, respectively, the entropy and fermion
number per unit volume.

For an ideal gas of Dirac neutrinos, the pressure is

\begin{equation}
P_{Dirac}=k_BT\Big\{2\int \Big({\frac{d^3p}{(2\pi \hbar )^3}}\Big) 
ln\Big(\big(1+e^{(\mu -\varepsilon )/k_BT}\big)
\big(1+e^{-(\mu +\varepsilon )/k_BT}\big)\Big)\Big\}.
\end{equation}

The point is that both the Dirac particle and antiparticle have a positive
energy $\epsilon $ but the fermion number is $+1$ for the particle and $-1$
for the antiparticle. Thus, the particle gets a chemical potential $\mu $
while the antiparticle gets a chemical potential $-\mu $.

Equations \ref{eq:relativity} and \ref{eq:grandpdirac} imply that

\begin{equation}
P_{Dirac}=\hbar c \Big({\frac{k_BT}{\hbar c}}\Big)^4 {\cal G}_{Dirac}\Big({%
\frac{mc^2}{k_BT}},{\frac{\mu }{k_BT}}\Big), 
\end{equation}

\noindent where

\begin{equation}
{\cal G}_{Dirac}(x,z)=\Big({\frac{1}{\pi^2}}\Big)\int_x^\infty \big(y\sqrt{%
y^2-x^2}\big)ln\Big(
\big(1+e^{(z-y)}\big)\big(1+e^{-(z+y)}\big)
\Big)dy. 
\end{equation}

In the zero mass limit 
\begin{equation}
\label{eq:diracpzero}\lim_{m\to 0}P_{Dirac}(T,\mu)=\hbar c \Big({\frac{k_BT}{%
\hbar c}}\Big)^4 {\cal G}\Big({\frac{\mu }{k_BT}}\Big), 
\end{equation}
\noindent where

\begin{equation}
{\cal G}(z)=\Big({\frac{1}{\pi^2}}\Big)\int_0^\infty y^2 ln\Big(\big(%
1+e^{(z-y)}\big)\big(1+e^{-(z+y)}\big)\Big)dy. 
\end{equation}

In the Fermi-Dirac degenerate regime 
\begin{equation}
P_o(\mu)=\lim_{T\to 0}\lim_{m\to 0}P_{Dirac}(T,\mu ), 
\end{equation}

\noindent one finds for the pressure and lepton number per unit volume, $%
n_o=(dP_o/d\mu )$,

\begin{equation}
P_o(\mu )=\Big({\frac{\hbar c}{12\pi^2}}\Big)
\Big({\frac{\mu }{\hbar c}}\Big)^4,\ \ \ n_o=\Big({\frac{1}{3\pi^2}}\Big)%
\Big({\frac{\mu }{\hbar c}}\Big)^3. 
\end{equation}

\noindent Thus, in the degenerate regime, 
\begin{equation}
\label{eq:degen}P_o=\Big({\frac{\hbar c}{12\pi^2}}\Big)\big(3\pi^2 |n_o|\big)%
^{4/3} \ \ \ (Dirac). 
\end{equation}

The chemical potential variations for the Dirac neutrino Eq.(\ref
{eq:diracpzero}) have no counterpart in the Majorana Eq.(\ref
{eq:majoranapzero}). In fact, in the general thermodynamic Eq.(\ref{eq:dp})
holds for Dirac neutrinos but must be restricted to $\mu _{Majorana}=0$ for
the Majorana case. Thus the equilibrium thermodynamic regime of Eq.(\ref
{eq:degen}) does not exist for Majorana neutrinos.

\section{Weak Interactions and the Approach to Equilibrium}

The arguments in the previous section are described in some mathematical
detail, but it is instructive to have a physical picture of what is
happening. Imagine the familiar filling up of energy levels with Dirac
fermions such as electrons. Each level can take two fermions in different
spin states. Now if the same procedure is attempted with the less-familiar
Majorana fermions, the two different spin states can now annihilate, leaving
that energy level vacant. This, physically, is the reason that there is no
Fermi energy is that there is no conserved lepton number and thus no Lagrange
multiplier (chemical potential) to enforce lepton conservation.

In the Standard Model, with the added hypothesis of Majorana neutrinos and
under normal conditions, the interactions of neutrinos with each other are
quite feeble, and the chance of a neutrino-number violating reaction is
rather small. In this case, true equilibrium may take a long time to be
achieved. One might then wish to consider an {\it approximately} conserved
quantity like the number of left-handed neutrinos less the number of
right-handed neutrinos, assuming then that left and right handed neutrinos
annihilate or are produced in pairs. An attendant Lagrange multiplier
analogous to chemical potential might be applied in the ideal gas case,
however, the corresponding equation of state will {\it not} correspond to 
{\it local thermal equilibrium}, and whether or not this is acceptable in a
given situation is a purely dynamical question.

There are situations in which neutrino densities are very high, and energy
densities are high enough that the weak interactions are effectively
unsuppressed. The difference in thermodynamic equations of state for Dirac
and Majorana neutrinos would then {\it clearly} have physical consequences.

\section{Application to Astrophysics and Cosmology}

While the main point of this paper is to point out the differences in the
thermodynamics of Dirac and Majorana fermions, we would also wish to explore
some physical systems where these ideas may be applied.

The thermodynamics of reactions in stars depends in large part on the nature
of the chemical potentials. Consider a reaction written as 
\begin{equation}
\sum_k z_k C_k\Leftrightarrow 0, 
\end{equation}

\noindent where $C_k$ is a particle component and $z_k$ is an integer
(positive or negative). A condition for thermal equilibrium is that

\begin{equation}
\sum_kz_k\mu _k=0,\ \ \ (equilibrium). 
\end{equation}
One simply replaces the symbolic component $C_k$ in the reaction with the
chemical potential $\mu _k$.

For example, thermal equilibrium for $e^++e^-\Leftrightarrow 2\gamma$,
implies $\mu_{e^+}+\mu_{e^-}=2\mu_\gamma $. Since $\mu_\gamma =0$, electrons
and positrons are in equilibrium if $\mu_{e^+}+\mu_{e^-}=0$. One may now add 
$e^++e^-\Leftrightarrow \nu +\bar{\nu}$, (Dirac), to derive the equilibrium
condition $\mu_{e^+}+\mu_{e^-}=\mu_\nu+\mu_{\bar{\nu}}$, (Dirac) so that $%
\mu_\nu+\mu_{\bar{\nu}}=0$, (Dirac). For the Majorana case we have (in close
analogy with photons), $e^++e^-\Leftrightarrow 2\nu $, (Majorana). This
requires $\mu_\nu =0$, (Majorana).

For weak nuclear reactions in stars, e.g. 
\begin{equation}
\label{eq:4.3}\nu +(Z,A)\Leftrightarrow (Z-1,A)+e^{-},\ \ \ 
\end{equation}
we have the equilibrium condition 
\begin{equation}
\mu _\nu +\mu _{(Z,A)}=\mu _{(Z-1,A)}+\mu _{e^{-}},\ \ \ (Dirac).
\end{equation}
\noindent Viewed as a Majorana neutrino reaction, Eq.(\ref{eq:4.3}) yields
the equilibrium condition 
\begin{equation}
\mu _{(Z,A)}=\mu _{(Z-1,A)}+\mu _{e^{-}},\ \ \ (Majorana).\ 
\end{equation}
\noindent Thus, with a nuclear chemical potential energy difference $%
\varepsilon ^{*}=\mu _{(Z,A)}-\mu _{(Z-1,A)}$ there will be a different
nuclear reaction equilibrium condition 
\begin{equation}
\varepsilon ^{*}=\mu _{e^{-}}-\mu _\nu \ \ (Dirac),\ \ \ \varepsilon
^{*}=\mu _{e^{-}}\ \ (Majorana).
\end{equation}
Eqs.(4,6) are equally true for the equilibrium of the beta decay reaction 
\begin{equation}
(Z,A)\Leftrightarrow (Z-1,A)+e^{-}+\bar \nu ,\ (Dirac),\ \ \
(Z,A)\Leftrightarrow (Z-1,A)+e^{-}+\nu ,\ (Majorana).
\end{equation}

Since the equilibrium distribution of nuclear particles is different in
models where the neutrinos are considered as Dirac particles or where the
neutrinos are considered as Majorana particles, it is evident that the
dynamics of the approach to equilibrium must also depend on whether the
neutrinos are considered as Dirac particles or as Majorana particles.

Of particular interest are supernovae, where most of the dynamics in the
core is driven by neutrinos and in fact, the neutrinos are assumed to be
degenerate \cite{ref:Sato}, a situation impossible for Majorana neutrinos in
local thermal equilibrium. Significant annihilation of neutrinos and
antineutrinos (or perhaps left and right handed Majorana neutrinos) has been
discussed by Goodman et al. \cite{ref:nunu}, indicating that there {\it are}
astrophysical systems in which there are significant (Standard Model)
neutrino-neutrino interactions. Estimates of effects of a neutrino number
non-conserving interactions in the context of supernova explosions and
equilibrium equations of states for Dirac and Majorana equations of state
would be of interest.

There are possible cosmological implications as well, where the process $\nu
+\bar \nu \rightarrow e^{+}+e^{-}$ is assumed to be bring the neutrinos to
equilibrium at temperatures greater than about $1\,MeV$ \cite{ref:Sciama}.
Indeed, it is possible that the neutrinos were, if Dirac, degenerate in the
early universe \cite{ref:Dolgov}, a situation impossible for Majorana
neutrinos in equilibrium. Shapiro et al. \cite{ref:Shapiro} have pointed out
difficulties in right-handed massive Dirac neutrinos coming to thermal
equilibrium in the early universe with only Standard Model interactions,
indicating again, that the approach to equilibrium and the approximations
are dynamical questions.

More exotic than neutrinos, but of great current interest, are
supersymmetric theories \cite{ref:SUSY}, which hypothesize a massive
Majorana partner for every observed boson. In particular, there should be
neutral particles such as partners to the photon, which are fermions and
which are their own antiparticles. These particles, in equilibrium, must
obey the Majorana equation of state as given above. At low energies they
must be quite massive to have avoided production and detection at
accelerator experiments, but in the early universe before the symmetry
breaking that gave them mass they were presumably massless Majorana fermions.

\section{Zero Chemical Potential for Dirac Fermions with Lepton Number
Violation}

We note in passing that it is not only the possibility that neutrinos are
Majorana that could force their chemical potential to be zero and generate a
Majorana-like equation of state. In fact, the presence of any interactions
violating lepton number conservation, such as those generically present in
grand-unified theories \cite{ref:GUTs} and, indeed, in the Standard Model
itself \cite{ref:tHooft}, are sufficient, if one is concerned about {\it %
local equilibrium} distributions. At some level, given the apparent
matter-antimatter asymmetry in the universe, it seems likely that some
lepton (and indeed baryon) number violation took place in the early universe.

\section{Conclusions}

We have pointed out that the thermodynamic equations of state for Majorana
and Dirac particles are quite different, even in the massless limit. This
difference is essentially due to the fact that Majorana fermions are their
own antiparticles and there is no conserved fermion number, no corresponding
chemical potential, and no possibility of forming a degenerate fermion gas
in local equilibrium for the Majorana case. We have noted that there is a
subtlety involved, since the electro-weak approach to equilibrium can be
quite slow if the interactions between fermions of the same helicities are
sufficiently small, and we have discussed the case of Dirac and Majorana
neutrinos in some detail. Possible applications have been explored.

\end{document}